\documentclass[reprint,aps,superscriptaddress,amsmath,amssymb, nofootinbib]{revtex4-2}

\usepackage{graphicx}
\usepackage{float}
\usepackage{physics}
\usepackage[utf8]{inputenc}
\usepackage{float}
\usepackage{hyperref}	
\usepackage{amsmath}

\begin{document}
	
	\title{Dimensional crossover for universal scaling far from equilibrium}
	\author{Lasse Gresista}
	\affiliation{Institut für Theoretische Physik, Universität Heidelberg, Philosophenweg 16, 69120 Heidelberg, Germany}
	\affiliation{Institute for Theoretical Physics, University of Cologne, 50937 Cologne, Germany}
	\author{Torsten V. Zache}
	\affiliation{Institut für Theoretische Physik, Universität Heidelberg, Philosophenweg 16, 69120 Heidelberg, Germany}
	\affiliation{Institute for Quantum Optics and Quantum Information of the Austrian Academy of Sciences, 6020 Innsbruck, Austria}
	\affiliation{Center for Quantum Physics, University of Innsbruck, 6020 Innsbruck, Austria}
	\author{Jürgen Berges}
	\affiliation{Institut für Theoretische Physik, Universität Heidelberg, Philosophenweg 16, 69120 Heidelberg, Germany}
	\begin{abstract}
		We perform a dynamical finite-size scaling analysis of a nonequilibrium Bose gas which is confined in the transverse plane.  Varying the transverse size, we establish a dimensional crossover for universal scaling properties far from equilibrium. Our results suggest that some aspects of the dynamical universal behavior of anisotropic systems can be classified in terms of fractional spatial dimensions. We discuss our findings in view of recent experimental results with quasi one-dimensional setups of quenched ultracold quantum gases.
	\end{abstract}
	\maketitle
	
	\section{Introduction}
	
	The classification of universal many-body phenomena far from equilibrium, similar to what has been successfully achieved for systems close to thermal equilibrium~\cite{RMP49}, is an outstanding problem in physics. In the past years, new forms of universal nonequilibrium behavior have been theoretically predicted~\cite{PRL101,PRL114}, which represent essential ingredients in our understanding of the (pre-)heating process in the early universe~\cite{PRD70} and the early stages of high-energy nuclear collisions~\cite{PRD89}. While originally proposed in the context of high-energy physics, the actual discovery of this dynamical universal phenomenon was first achieved in table-top experiments with ultracold quantum gases~\cite{ober18,schmied18,hadzi20}. These developments across subdisciplines trigger important progress for a wide range of applications, see e.g.\ Refs.~\cite{orioli15,PRA99,IMP34,PRL125} and \cite{RMP21} for a review.
	
	For equilibrium universality classes it is well-established that only general properties matter, such as symmetries of the physical system, rather than microscopic details. In particular, universal quantities like scaling exponents characterizing critical phenomena in equilibrium depend on the dimensionality of space~$d$~\cite{CUP95}. Much less is known about the dependence of universal scaling properties far from equilibrium on the dimensionality or geometry of space. This can be of considerable importance for the interpretation of experimental results, such as obtained from quenched ultracold quantum gases in confined geometries. Corresponding experiments were performed in quasi one-dimensional traps~\cite{ober18,schmied18}, where the longitudinal direction $L_L$ is much larger than transverse trap sizes $L_T$, while more recent results on far-from-equilibrium scaling concern homogeneous Bose gases with $L_T \simeq L_L$~\cite{hadzi20}. The different setups seem to indicate a rather strong dependence of the scaling properties on the effective dimensionality of space, especially close to one spatial dimension emerging from $L_T \ll L_L$. 
	
	In this work, we compute the universal scaling properties of a dilute Bose gas far from equilibrium, whose properties are described by a non-relativistic theory for a complex scalar field. Starting from a two-dimensional geometry with equal longitudinal and transverse sizes, $L_L = L_T$, we decrease $L_T$ successively until an effectively one-dimensional behavior is observed. We perform a dynamical finite-size scaling analysis to extract numerically the universal scaling exponents and scaling functions as a function of $L_T$, which represent the main results of our study. Since the far-from-equilibrium systems exhibit a self-similar spatio-temporal evolution, our dynamical finite-size analysis involves comparisons of systems with various spatial sizes at given different times.  Remarkably, several characteristic quantities exhibit an apparently smooth crossover from two to one spatial dimension. Our results may help to shed some light on the experimental findings for the different quasi one-dimensional setups of Ref.~\cite{schmied18} showing much smaller scaling exponents than Ref.~\cite{ober18}. 
	
	\section{Dimensional crossover on a lattice}
	Before presenting our main results, we briefly introduce the model and our main observables in this section. In subsection \ref{sec:Initialconditions}, we set the stage for the dimensional crossover by identifying three qualitatively distinct parameter regimes that are analyzed in detail in the following sections \ref{sec:2D} and \ref{sec:crossover}.
	
	\subsection{Non-relativistic scalar field theory}
	\label{sec:lattice}
	
	The dynamics of a dilute Bose gas may be described by a non-relativistic scalar field theory. We consider a complex valued field $\Psi(t,\textbf{x})$ characterized by the Gross-Pitaevskii equation as the classical equation of motion,
	\begin{equation}
		\label{eq:gpe}
		i\partial_t\Psi(t,\textbf{x})=\qty(-\frac{\nabla^2}{2m}+g\Psi^\dagger(t,\textbf{x})\Psi(t,\textbf{x}))\Psi(t,\textbf{x}),
	\end{equation}
	where $t$ and $\textbf{x}$ denote the time and space variables, $g$ is the coupling constant, and $m$ is the mass of the gas particles.
	
	For an arbitrary operator $O[\Psi,\Psi^\dagger]$, observables are defined for any given density matrix $\rho_0$ specified at some initial time $t_0$ as $\expval{O[\Psi,\Psi^\dagger]}\equiv\Trace{\qty{\rho_0 O[\Psi,\Psi^\dagger]}}$. In a quantum field theory the trace gives the expectation value of the quantity, while in a classical-statistical field theory it represents the ensemble average. Here we focus on correlation functions of fields, such as the `statistical' two-point correlation function
	\begin{equation}
		\label{eq:statisticaltwopoint}
		F(t,\textbf{x}-\textbf{y})=\expval{\frac{1}{2}\qty{\Psi(t,\textbf{x})\Psi^\dagger(t,\textbf{y})+\Psi^\dagger(t,\textbf{y})\Psi(t,\textbf{x})}}
	\end{equation}
	for spatially homogeneous systems. 
	
	We evaluate the fields on a two-dimensional spatial lattice with periodic boundary conditions, providing a well-defined setup for a finite-size scaling analysis. The physical longitudinal/transverse momenta $p_{L/T, n}$ on the lattice are then obtained from the eigenvalues of the discretized form of the negative Laplace operator ($-\nabla^2$) as
	\begin{equation}
		p_{L/T, n}^2 = \frac{4}{a_S^2} \sin^2 \qty(\frac{\pi n}{N_{L/T}}),
		\label{eq:momenta}
	\end{equation}
	where $n \in \qty{0,...,N_{L/T}-1}$ with $N_{L/T}$ denoting the number of lattice points in the longitudinal/transverse direction with lattice spacing $a_s$. Apart from the case $L_T = L_L$, we will consider mainly systems with $L_T<L_L$ and for small enough transverse lattice lengths finite-size effects are expected to become relevant for the computation of observables. We set up the simulations with $L_L$ much larger than all characteristic length scales of the theory, such that we can essentially neglect longitudinal finite-size effects. As a consequence, observables in momentum space can be approximately taken to depend on continuous longitudinal momenta whereas the discrete nature of the transverse momenta can play an important role. 
	
	\subsection{Energy density and initial conditions}
	\label{sec:Initialconditions}
	
	In our setup, for the Gross-Pitaevskii field theory described by Eq.~\eqref{eq:gpe} we can estimate the (conserved) energy for Gaussian initial conditions at time $t_0$ as
	\begin{eqnarray}
		\langle H \rangle &\approx& \int \mathrm{d} x_L \int_0^{L_T} \!\! \mathrm{d} x_T \Big\{ g \langle \Psi^\dagger(t_0,x_L,x_T)   \Psi(t_0,x_L,x_T)  \rangle^2 \nonumber\\
		&-& \frac{1}{2m} \langle \Psi^\dagger(t_0,x_L,x_T) \left( \partial_L^2 + \partial_T^2 \right)  \Psi(t_0,x_L,x_T)  \rangle 
		\Big\} .
		\label{eq:Hx}
	\end{eqnarray}
	For the spatially translation invariant system in Fourier space, with
	\begin{eqnarray}
		&& \langle \Psi^\dagger(t_0,x_L,x_T)   \Psi(t_0,x_L,x_T)  \rangle = \nonumber\\
		&& \int \frac{\mathrm{d}p_L}{2\pi} 
		\frac{1}{L_T} \sum_n \langle \Psi^\dagger_n   \Psi_n  \rangle (t_0,p_L) ,
	\end{eqnarray}
	the energy density $\epsilon \equiv \langle H \rangle/(L_L L_T)$ is given by a sum over discrete momentum modes $n$:
	\begin{eqnarray}
		\epsilon &=& \int \frac{\mathrm{d}p_L}{2\pi} \frac{1}{L_T} \sum_n \frac{p_L^2 + (2\pi n/L_T)^2}{2m} 
		\langle \Psi^\dagger_n   \Psi_n  \rangle (t_0,p_L) \nonumber\\
		&+&  g \left( \int \frac{\mathrm{d}p_L}{2\pi} \frac{1}{L_T} \sum_n \langle \Psi^\dagger_n   \Psi_n  \rangle (t_0,p_L)\right)^2 \nonumber\\
		&=& \frac{\epsilon^{1{\mathrm D}}}{L_T} +  \frac{\epsilon^{\mathrm{ex}}}{L_T}  .
		\label{eq:Hp}
	\end{eqnarray}
	In the last decomposition we identify with $\epsilon^{1{\mathrm D}}$ the energy density of a one-dimensional field theory for the zero mode $\Psi_0$, i.e.\
	\begin{eqnarray}
		\epsilon^{1{\mathrm D}} &=& \int \frac{\mathrm{d}p_L}{2\pi} \frac{p_L^2}{2m}  \langle \Psi^\dagger_0   \Psi_0  \rangle (t_0,p_L) \nonumber\\
		&+& g^{1{\mathrm D}} \left( \int \frac{\mathrm{d}p_L}{2\pi} \langle \Psi^\dagger_0   \Psi_0  \rangle (t_0,p_L)\right)^2,
		\label{eq:epsilon1D}
	\end{eqnarray}
	where we introduced the effective coupling $g^{1{\mathrm D}} \equiv g/L_T$, whose strength depends on the transverse lattice size. The rest of the original energy density is taken into account in
	\begin{eqnarray}
		&&\epsilon^{\mathrm{ex}} = \int \frac{\mathrm{d}p_L}{2\pi} \frac{p_L^2+ (2\pi/L_T)^2}{2m}  \langle \Psi^\dagger_1   \Psi_1  \rangle (t_0,p_L) \nonumber\\
		&&+\, 2 g^{1{\mathrm D}} \int \frac{\mathrm{d}p_L}{2\pi} \langle \Psi^\dagger_0   \Psi_0  \rangle (t_0,p_L)
		\int \frac{\mathrm{d}k_L}{2\pi} \langle \Psi^\dagger_1   \Psi_1  \rangle (t_0,k_L) \nonumber\\
		&&+\, g^{1{\mathrm D}} \left( \int \frac{\mathrm{d}p_L}{2\pi} \langle \Psi^\dagger_1   \Psi_1  \rangle (t_0,p_L)\right)^2
		\nonumber\\
		&&+ \left(\mbox{terms involving also modes $\Psi_n$ with} \,\, n \ge 2 \right) \, ,
	\end{eqnarray}
	which contains the interactions with all other modes. One observes that fields $\Psi_n $ with $n \neq 0$ exhibit an effective `chemical potential' $\mu^{1{\mathrm D}}_n =  -(2\pi n/L_T)^2/(2m)$. As a consequence, their excitation becomes energetically costly for sufficiently small transverse size $L_T$. 
	
	We consider a class of Gaussian initial states that are fully characterized by a vanishing field expectation value $\langle \Psi_n \rangle = \langle \Psi^\dagger_n \rangle = 0$ and the correlation\footnote[1]{More precisely, for small transverse sizes the term $2\pi n/L_T$ has to be replaced by $p_{T,n}$ in this equation.}. 
	\begin{eqnarray}
		\langle \Psi^\dagger_n   \Psi_n  \rangle (t_0,p_L) &=& \frac{A}{m g}\, \Theta\Big( Q - \sqrt{p_L^2 + (2\pi n/L_T)^2}\Big) 
		\nonumber\\
		&+& \frac{1}{2}\, \Theta\Big( \Lambda - \sqrt{p_L^2 + (2\pi n/L_T)^2}\Big). 
		\label{eq:initialcond}
	\end{eqnarray}
	Using the Heaviside step function $\Theta$, this initial correlation is composed of excitations up to a characteristic momentum scale $Q$ and a vacuum `quantum-half' up to an ultraviolet scale $\Lambda$ with $\Lambda \gg Q$. Again, $L_L$ is always taken to be much larger than all characteristic length scales of the theory such that $L_L \gg 2 \pi/Q$. We also ensure that $\Lambda$ never exceeds the highest possible lattice momenta related to the inverse lattice spacing $a_s^{-1}$.
	
	The amplitude of the excitations will be taken to be much larger than unity, $A/(mg) \gg 1$, such that the system is far from equilibrium initially. For comparison, excitations of a thermal state exhibit at a characteristic temperature amplitudes or `occupancies' of order unity. For given $\Lambda$ we choose the amplitude $A/(mg)$ large enough such that the initial energy density is dominated by contributions up to the momentum scale $Q$. In this highly occupied case, the time evolution of the quantum theory may be approximated by the corresponding classical-statistical field theory dynamics while keeping the same initial conditions for correlation functions as in the original quantum theory~\cite{RMP21}. In the following we adopt the classical-statistical approximation also for the lattice setup we consider.     
	
	By changing the transverse extension $L_T$, with Eq.~\eqref{eq:initialcond} we change the initial condition for given other parameters. For instance, if both the longitudinal and transverse lattice sizes are much larger than the characteristic inverse momentum scale, i.e.\ also $L_T \gg 2\pi/Q$, then all the corresponding modes of the two-dimensional field theory are excited initially. By contrast, for $L_T \ll 2\pi/Q$ only the zero mode with $n=0$ is excited, while all other modes with $n \ge 1$ up to the ultraviolet cutoff are in vacuum with amplitude given by the quantum-half. Finally, once $L_T \ll 2\pi/\Lambda$ only the zero mode and no other modes are excited initially\footnote[2]{To be precise, whether or not modes other than the zero mode are initially excited for a given $L_T$ depends on the value of the lowest non-zero transverse momentum $p_{T, 1}$  given by Eq.~\eqref{eq:momenta} with $n = 1$. For $p_{T, 1} < Q$ modes with $n\geq1$ up to $Q$ are excited, for $Q < p_{T, 1} < \Lambda$ only the zero mode is excited while modes with $n>1$ up to $\Lambda$ are in vacuum and for $\Lambda < p_{T, 1} $ the initial amplitude of all modes with $n\geq1$ is exactly zero.}. 
	This last scenario is not realized in a realistic quantum field theory since $L_T$ cannot be sensefully made smaller than $a_s$. However, we keep this possibility by also artificially separating the scales $\Lambda$ and $a_s^{-1}$ in order to demonstrate that a full dimensional reduction in the dynamics is observed in our setup only in the absence of initial vacuum fluctuations. Therefore, from the initial distribution of the energy density across modes we distinguish three parameter regimes, which may be expected to lead to different subsequent dynamics:
	\begin{enumerate}
		\item $L_T \gg 2\pi/Q$: two-dimensional dynamics,
		\item $2\pi/\Lambda \ll L_T \ll 2\pi/Q$: effective dynamics for highly occupied zero mode modified by interactions with higher (vacuum) modes,
		\item $L_T \ll 2\pi/\Lambda$: one-dimensional dynamics for zero mode, no other modes excited.
	\end{enumerate}
	As the system evolves in time for $t > t_0$, different scales can arise dynamically, which was previously demonstrated for the isotropic case~\cite{orioli15}. In the next section we first show that for $L_T \gg 2\pi/Q$ we recover in our setup the self-similar scaling results expected for the two-dimensional field theory. In Sec.~\ref{sec:crossover} we then establish the crossover from two to one spatial dimensions via the parameter regimes 2 and 3.
	
	\section{Self-similar scaling in two dimensions}
	\label{sec:2D}
	
	\begin{figure}[t]
		\includegraphics[width=\columnwidth]{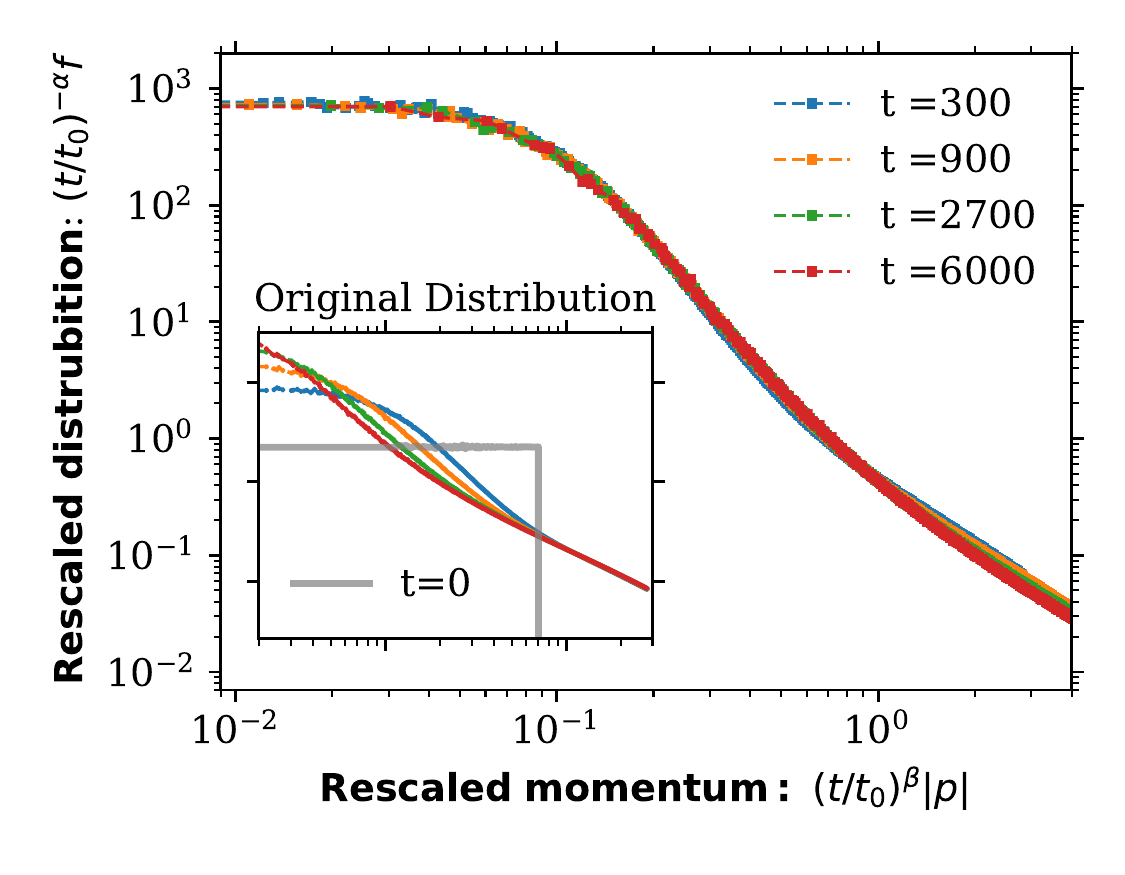}
		\caption{Rescaled distribution as a function of the rescaled momentum in two spatial dimensions with occupation exponent $\alpha=1.032$, correlation exponent $\beta=0.536$ (see Fig.~\ref{fig: exponents_isotropic}) and reference time $t_0 = 300$. The inset shows the time evolution without rescaling. The rescaled curves collapse to the time independent scaling function $f_S$.}
		\label{fig: distribution_isotropic}
	\end{figure}
	
	To characterize the behavior of our nonequilibrium system in two spatial dimensions, we employ the statistical two-point correlation [Eq.~\eqref{eq:statisticaltwopoint}]. This allows us to define a distribution function $f(t,\textbf{p})$ in momentum space via 
	\begin{equation}
		f(t,\textbf{p})+\frac{1}{2}=\int \mathrm{d}^2x \, e^{-i\textbf{px}}F(t,\textbf{x}).
		\label{eq:distribution}
	\end{equation} 
	The vacuum quantum-half on the LHS of this definition is identified with vanishing particle number, $f=0$, and the total particle number is conserved: $\int \mathrm{d}^2 p\, f(t,\textbf{p}) = \text{const}$. 
	
	To compute the nonequilibrium time evolution, we repeatedly solve the classical field equation \eqref{eq:gpe} numerically on a spatial lattice as described in Sec.~\ref{sec:lattice}. To evolve in time we use a split-step algorithm \cite{javanainen04} and observables are obtained from averaging the results over an ensemble of initial field configurations sampled from a Gaussian distribution~\cite{orioli15}. The number of runs is increased until convergence of results is observed. We employ a square lattice with $L_L= L_T=1024$, where all quantities are given in units of appropriate powers of the lattice spacing $a_s$. We checked that with this choice finite-size effects can be neglected. The highly occupied initial conditions specified by Eq.~\eqref{eq:initialcond} of Sec.~\ref{sec:Initialconditions} are realized with an amplitude $A=25/(mg)$. For the further parameters of the simulation we choose the characteristic momentum to be $Q=0.5$, the coupling as $g=1$ and the mass as $m=0.5$. We emphasize that none of the universal properties we are going to extract depend on the detailed choices of these parameters. Moreover, because of the high occupancies of modes in this two-dimensional setup the quantum-half in the initial condition can be neglected to very good accuracy in this case. 
	
	Starting from the initial distribution function 
	\begin{equation}
		f(t_0=0, \textbf{p})= \frac{A}{mg} \, \Theta(Q-\abs{\textbf{p}}),
		\label{eq: initial conditions infinite}
	\end{equation} 
	the time evolution of $f(t, \textbf{p})$ is illustrated in Fig.~\ref{fig: distribution_isotropic}. The inset shows the distribution at different snapshots in time as a function of the modulus of spatial momentum $\abs{\textbf{p}}$. The main figure shows the same data but rescaled as $(t/t_0)^\alpha f(t, \textbf{p})$ as a function of the rescaled momentum $(t/t_0)^\beta\abs{\textbf{p}}$, where the reference time $t_0$ is the earliest time shown. The curves at different times collapse to a time-independent scaling function $f_S$ defined via
	\begin{equation}
		f(t, \textbf{p}) = (t/t_0)^{\alpha} f_S\left((t/t_0)^\beta\abs{\textbf{p}}\right)
		\label{eq:scaling}
	\end{equation}
	after the rescaling with the `occupation' exponent  $\alpha$ and the `correlation' exponent $\beta$. This self-similar scaling behavior is associated with nonthermal infrared fixed points whose universal low-momentum properties have been established previousl~\cite{orioli15}. A relation between $\alpha$ and $\beta$ can be obtained by imposing particle number conservation,
	\begin{equation}
		\int d^dp \,f(t,\textbf{p})=(t/t_0)^{\alpha-\beta d}\int d^dq \, f_S(\textbf{q})=\text{const},
		\label{eq:numberconservation}
	\end{equation} 
	such that $\alpha=\beta d$ with spatial dimension $d$.
	
	\begin{figure}[t]
		\includegraphics[width=\columnwidth]{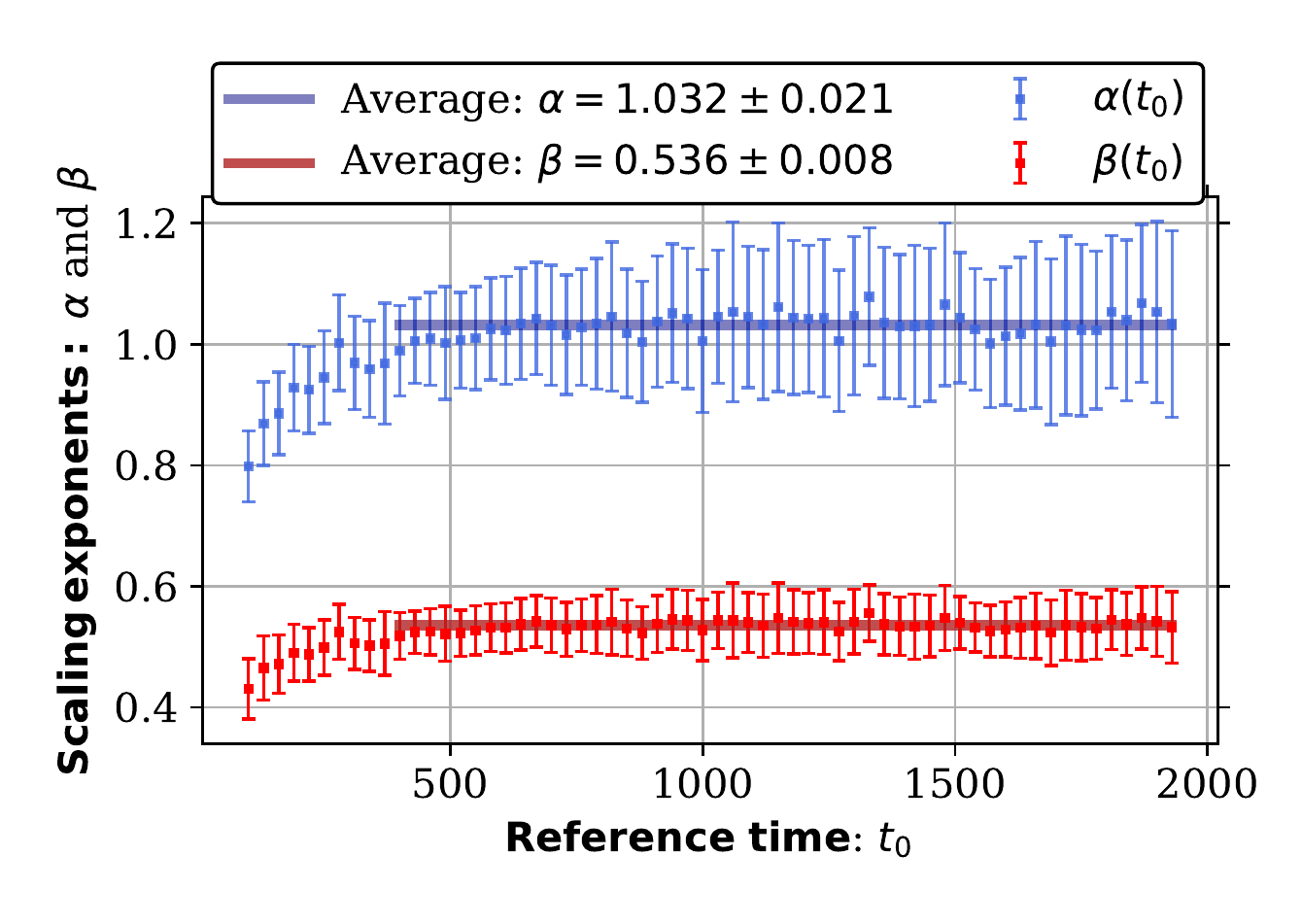}
		\caption{Occupation and correlation exponents for different values of the reference time $t_0$ for two spatial dimensions. The solid lines depict the average exponents and the reference times included in the averaging. The exponents are seen to become insensitive to the choice of the reference time after the onset of self-similar dynamics.}
		\label{fig: exponents_isotropic}
	\end{figure}	
	
	Here the scaling exponents $\alpha$ and $\beta$ are obtained through a fitting procedure which considers a reference time $t_0$ together with two additional times $t_1=2t_0$ and $t_2=3t_0$ such that the exponents minimize the difference of the corresponding rescaled curves. This is done for many different reference times where self-similarity is observable and the values of the scaling exponents are obtained from averaging. Fig.~\ref{fig: exponents_isotropic} shows the scaling exponents as a function of the reference time, where the solid lines illustrate the average values as well as the reference times included in the average. While during the initial evolution there is a redistribution of modes, very quickly self-similar scaling sets in and the exponents become insensitive to the reference time chosen. We find $\alpha=1.032\pm0.021$ and $\beta=0.536\pm0.008$ consistent with previous estimates~\cite{orioli15}. 
	
	To give uncertainties for the average exponents we calculate the 16\% and 84\% percentile from all the values included in the averaging. As the lower error bound we give the difference of the average from the 16\% and for the upper bound the difference from the 84\% percentile, which in this case turn out to be symmetric. The errorbars in Fig.~\ref{fig: exponents_isotropic} for the scaling exponents for a specific reference time are estimated errors from the fitting procedure (for details see appendix \ref{appendix: fitting procedure}).
	
	\section{Crossover from two dimensions to one dimension}
	\label{sec:crossover}
	
	\begin{figure*}[t]
		\includegraphics[width=\columnwidth]{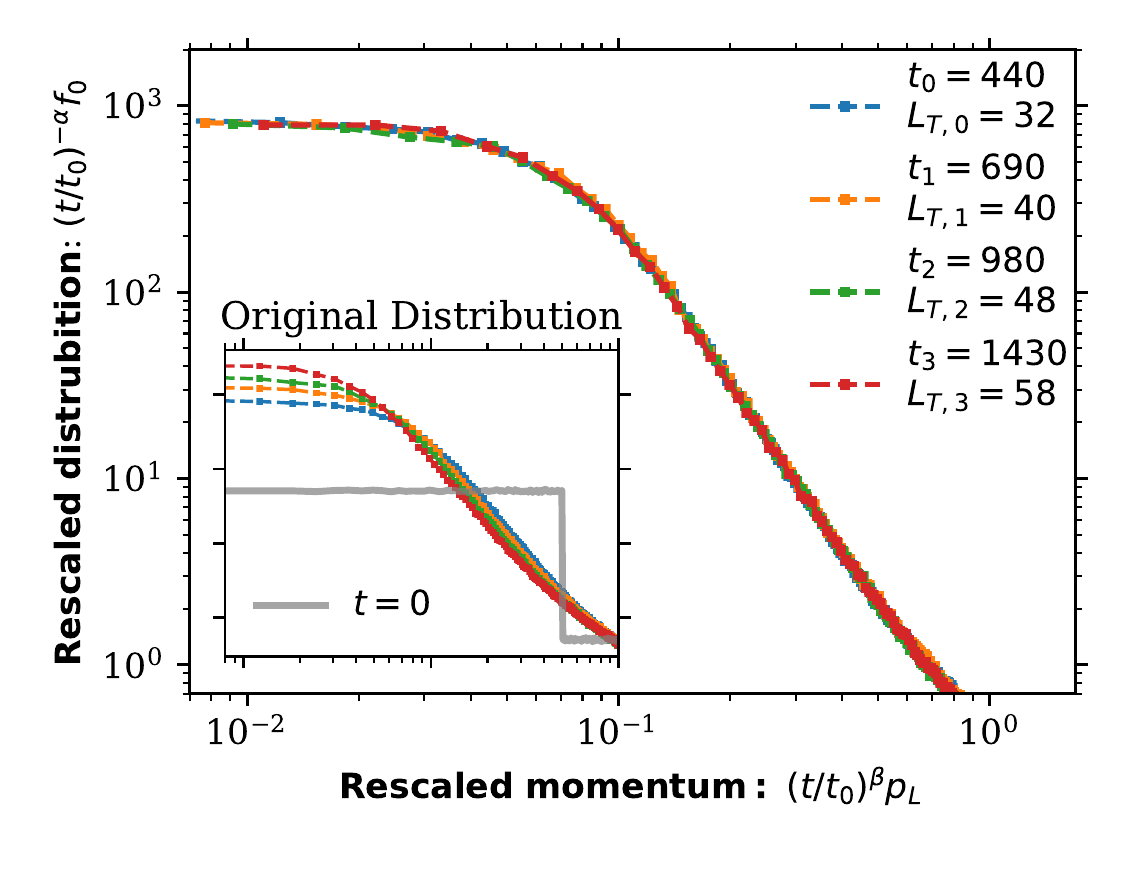}
		\includegraphics[width=\columnwidth]{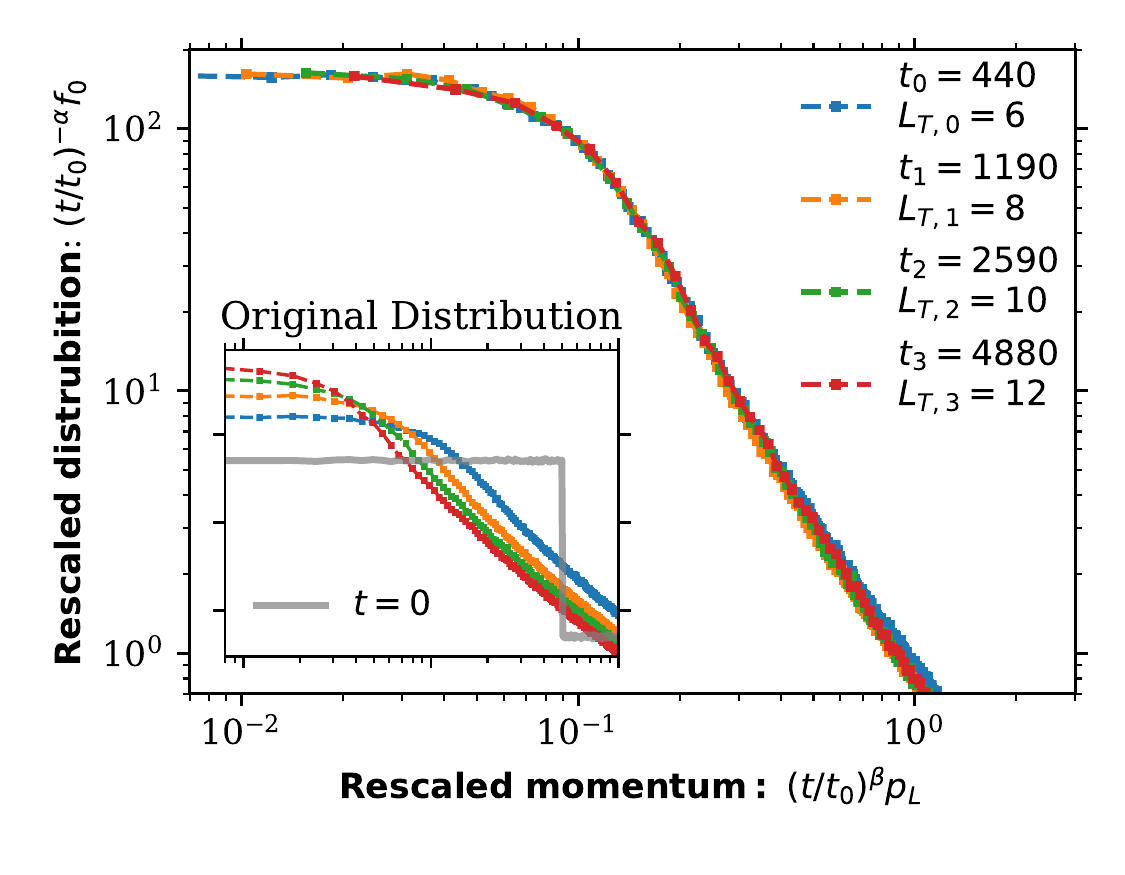}
		\vspace{-0.5cm}
		\caption{Rescaled distribution $f_0$ as a function of the rescaled momentum for different transverse lattice sizes. The  various $L_{T,i}$ at different times $t_i$ reflect the dynamical finite-size scaling. The inset shows the original curves without rescaling and the initial distribution. After rescaling all curves lie well on top of each other, displaying the scaling function $f_{0,S}$. For the left plot ($L_{T,0}=32$) the scaling exponents are $\alpha=0.96$, $\beta=0.50$, $\zeta=0.50$ and for the right plot ($L_{T,0}=6$) they are $\alpha=0.55$, $\beta=0.52$, $\zeta=0.29$.}
		\label{fig: distribution finite-size}
	\end{figure*}
	
	\subsection{Dynamical finite-size scaling analysis}
	Starting from the two-dimensional setup described in the last section, we now consider simulations with decreasing transverse lattice size until the system becomes essentially one-dimensional. For small transverse lattices finite-size effects are expected to become relevant and the correlation functions will in general depend explicitly on $L_T$. Following Eq.~\eqref{eq:distribution} we define the mode resolved distribution functions
	\begin{eqnarray}
		&& f_n(t,p_L,L_T)+\frac{1}{2}= \nonumber\\
		&&\int \mathrm{d} x_L\, e^{-i p_L x_L} \int_0^{L_T} \!\! \mathrm{d} x_T\, e^{-i 2\pi n x_T/L_T } F(t,x_L,x_T). \quad
		\label{eq:distribution2}
	\end{eqnarray}
	Here we are interested in the low-frequency behavior and study the scaling of the correlator zero-mode $f_0(t,p_L,L_T)$ for longitudinal infrared momenta $p_L$. Similar to Eq.~\eqref{eq:scaling}, a self-similar regime is observed if
	\begin{equation}
		f_0(t, p_L , L_T) = (t/t_0)^{\alpha} f_{0,S}\left((t/t_0)^\beta p_L,(t/t_0)^{-\zeta} L_T\right)
		\label{eq:finitesizescaling}
	\end{equation}
	where the additional exponent $\zeta$ describes the finite-size scaling of the system. As a consequence, in order to establish scaling for different times we have to consider systems with various transverse sizes. For this purpose we choose a reference time $t_0$ in the evolution of a system with transverse size $L_{T, 0}$ and compare it to some other time $t_i$ in the evolution of a system with transverse size 
	\begin{equation}
		L_{T,i} = (t_i/t_0)^{\zeta} L_{T, 0}\, .
		\label{eq: times_lengths_relation}
	\end{equation} 
	According to Eq.~\eqref{eq:finitesizescaling} we can then extract a possible scaling function $f_{0,S}$ for any given size $L_{T,0}$ by plotting $(t_i/t_0)^{-\alpha} f_0(t_i, p_L , (t_i/t_0)^{\zeta} L_{T, 0})$ as a function of $(t_i/t_0)^\beta p_L$. For this purpose we choose four different lengths $L_{T,i}$ ($i=0,1,2,3$) and the corresponding times $t_i$ are then calculated from Eq.~\eqref{eq: times_lengths_relation}. To obtain the scaling exponents we apply the same fitting procedure as for the two-dimensional case, but now additionally also vary $\zeta$ and select the best fit value. From Eq.~\eqref{eq: times_lengths_relation} it can be seen that for different given lengths $L_i$ increasing the value of $\zeta$ lowers the differences between the times $t_i$. For large values of $\zeta$ the fit therefore considers curves that lie very close to each other such that the fit becomes less precise. However, in practice it turns out to be sufficient to limit the upper bound on the range of possible $\zeta$ values in order to find acceptable fits. While this works well for not too small transverse lattice sizes, it becomes less accurate when decreasing $L_{T,0}$ such that rather large uncertainties in the estimates for exponents can arise. The final estimates for the scaling exponents and their uncertainties are obtained by using multiple reference times and then calculating averages in the same way as for the isotropic two-dimensional system in Sec.~\ref{sec:2D}. Exemplifying figures showing the scaling exponents as a function of the reference time can be found in appendix~\ref{appendix: exponents time crossover}.
	
	\subsection{Dimensional crossover exponents and scaling functions}
	\label{sec: results exponents}
	
	We employ the initial conditions given in Eq.~\eqref{eq:initialcond} with the same parameter values as for the two-dimensional setup of Sec.~\ref{sec:2D} except that now we vary the transverse size as $L_T = 1024$, $512$, $256$, $128$, $64$, $58$, $48$, $40$, $32$, $26$, $22$, $18$, $16$, $14$, $12$, $10$, $8$, $6$, $4$, $2$, $1$. In contrast to the two-dimensional case, for small transverse lattices it is also crucial to include the quantum-half in the initial condition [see Eq.~\eqref{eq:initialcond}]. For $L_T \ll 2 \pi / Q$ only the zero mode is highly occupied initially while the higher modes are initialized with their vacuum amplitude. Moreover, for $L_T \ll 2 \pi / \Lambda$ no higher modes are initialized such that their classical dynamics becomes trivial and the highly occupied zero-mode is governed by a one-dimensional theory with the limitations as described in Sec.~\ref{sec:Initialconditions} (we choose $\Lambda=\sqrt{2.01}$).
	
	Fig.~\ref{fig: distribution finite-size} shows the rescaled distribution function as a function of the rescaled momentum at four different times for $L_{T,0}=32$ and $L_{T,0}=6$. For each time the distribution from a different $L_T$ is used, which captures the effect of finite-size scaling. The inset shows the original curves before rescaling, as well as the initial distribution. After rescaling the curves for different times all collapse very well to one curve, which is the scaling function of the system.
	
	\begin{figure}[t]
		\centering
		\includegraphics[width=\columnwidth]{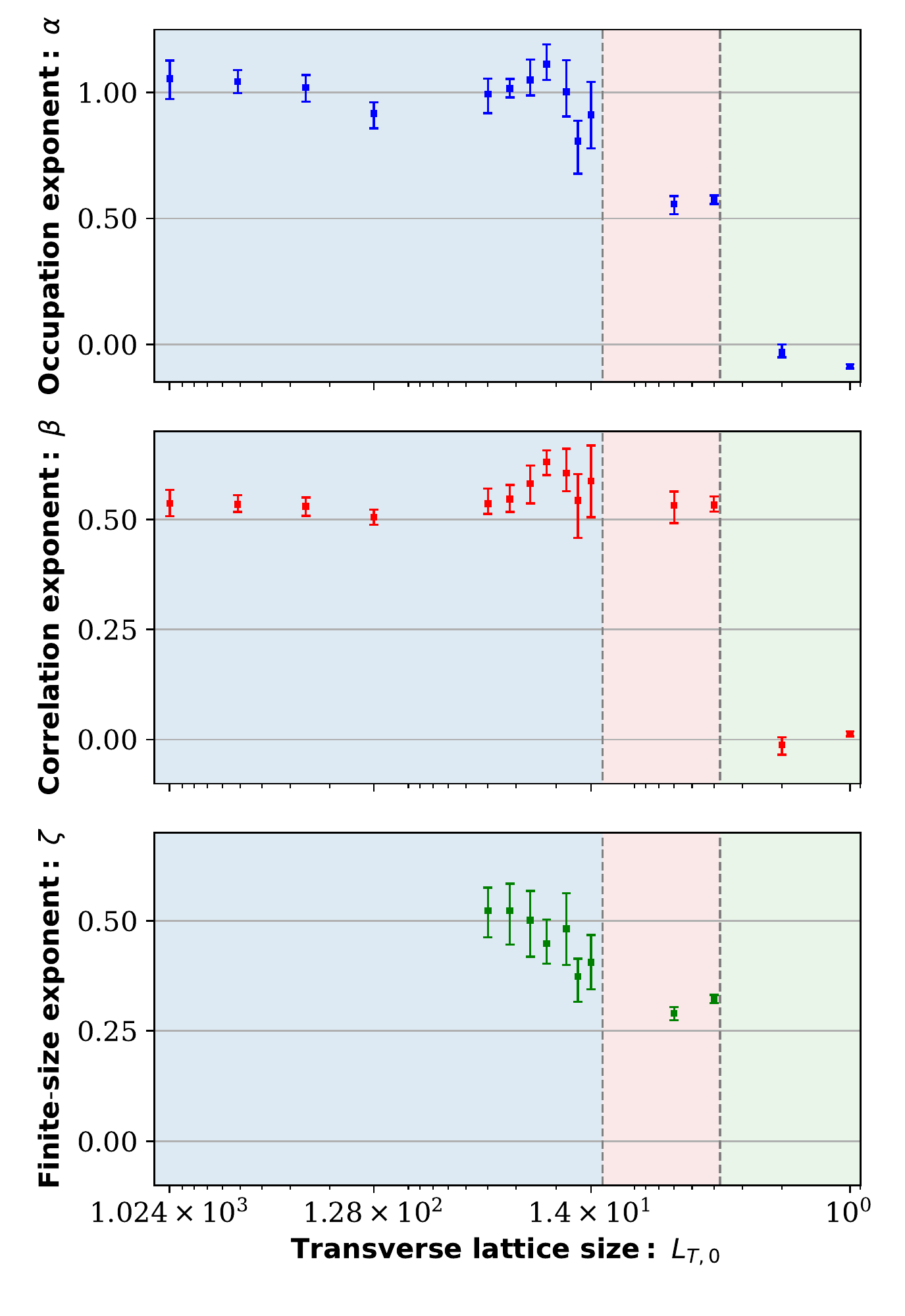}
		\vspace{-0.3cm}
		\caption{Scaling exponents as a function of the transverse lattice size $L_{T,0}$, which is the smallest lattice length considered in the corresponding fitting procedure. The exponent $\zeta$ is associated to finite-size scaling, which is only computed in the relevant regime as described in the main text. The dashed lines indicate the boundaries between the three parameter regimes at $L_T \approx 2\pi/Q$ and $L_T \approx 2\pi/\Lambda$.}
		\label{fig: exponents final}
	\end{figure}
	
	The average exponents as a function of $L_{T,0}$ are shown in Fig.~\ref{fig: exponents final}, which is the central result of this work. Here $L_{T,0}$ refers to the lowest transverse lattice size considered in the fitting procedure when finite-size scaling is relevant ($4\leq L_T\leq 64$), or to the plain transverse lattice size when no finite-size scaling is necessary and in which case $\zeta$ is not computed. We note that $\zeta$ approaches $\beta$ as $L_T \rightarrow L_L$, which is expected from dimensional analysis for an isotropic scaling behavior. The three background colors shown in the figure illustrate the three different parameter regimes for the initial conditions as explained in Sec.~\ref{sec:Initialconditions}. The scaling exponents $\alpha$ and $\zeta$ decrease significantly when decreasing $L_T$, while $\beta$ stays relatively constant for not too small $L_T$. We find the strongest decrease when transitioning from one parameter regime to another. In the green regime the dynamics for the zero mode becomes effectively one-dimensional. In this case we find exponents close to zero as expected for our class of initial conditions since, e.g., number conserving two-to-two scatterings are ineffective in one spatial dimension because of phase-space restrictions from energy and momentum conservation. 
	
\begin{figure}[h]
		\includegraphics[width=\columnwidth]{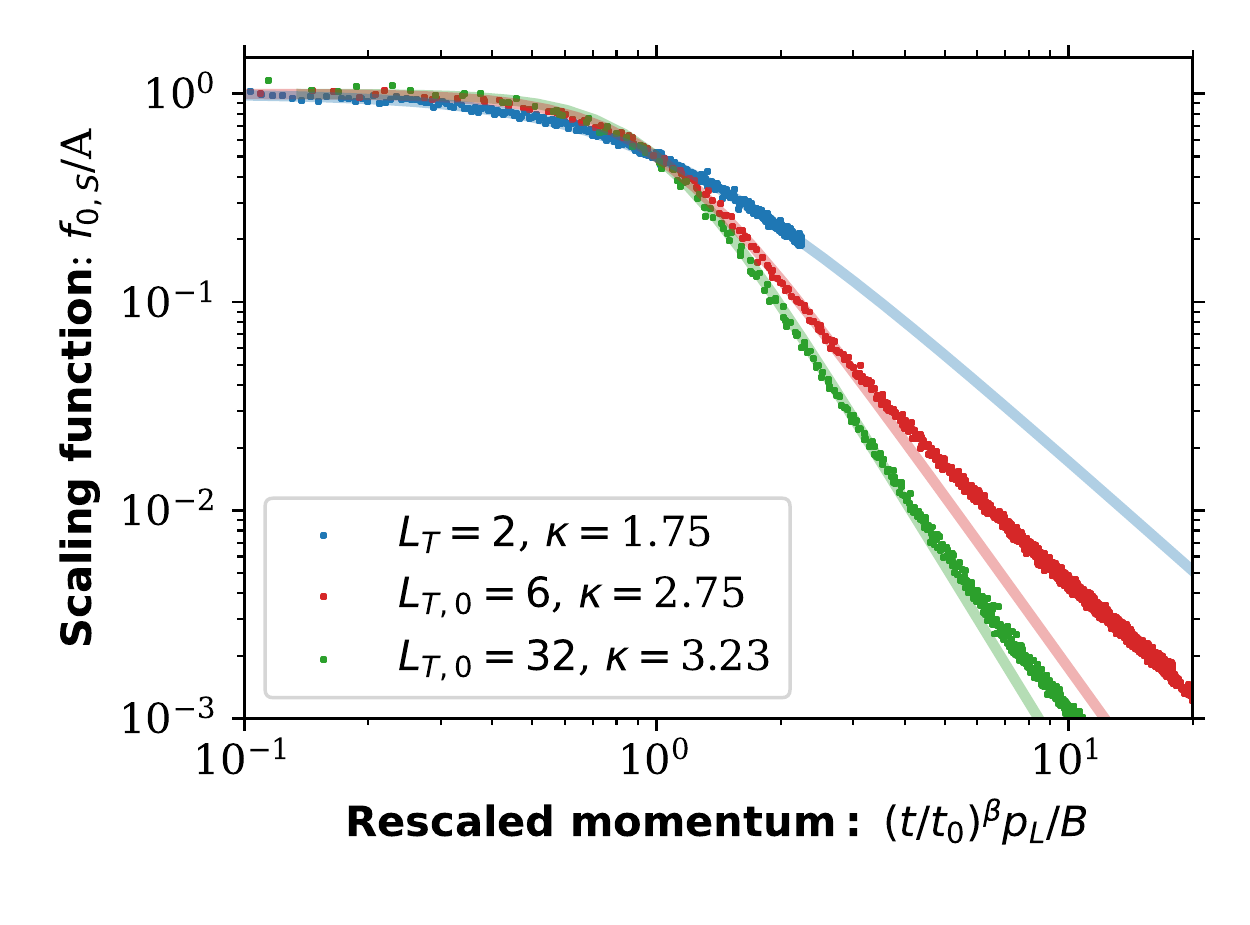}
		\vspace{-0.5cm}
		\caption{Fixed-point distribution $f_{0,S}$ for different $L_{T,0}$. The solid lines represent the fit of the function $f_S(\xi)=A/(1+(\xi/B)^\kappa)$, and the data is normalized by the amplitudes A and B to allow comparison between the regimes. The power-law exponent changes significantly with the transverse lattice size.}
		\label{fig: fs comparison}
	\end{figure}

	As can already be seen from Fig.~\ref{fig: distribution finite-size}, additionally to the scaling exponents the scaling function itself can change when decreasing $L_T$. For the relevant low-momentum range the scaling function is well described by the fit  \begin{equation}
		\label{eq: fit func}
			f_{0,S}(p)=A/\left(1+(p/B)^\kappa\right).
	\end{equation} 
	Here $\kappa$ is the `scaling function' exponent and the amplitudes $A$ and $B$ are two non-universal constants, which we have to extract from the data in order to compare the distributions of different systems. This is done in Fig.~\ref{fig: fs comparison} which shows the normalized scaling function as a function of the rescaled momentum in the three different regimes. The solid lines depict the corresponding fits. For the curve of $L_T=2$ no rescaling was performed since in this regime we found $\alpha \approx \beta \approx 0$. We observe that the exponent $\kappa$ changes between the three regimes, while the overall shape of the curve remains relatively similar.
	
\begin{figure}[h]
		\includegraphics[width=\columnwidth]{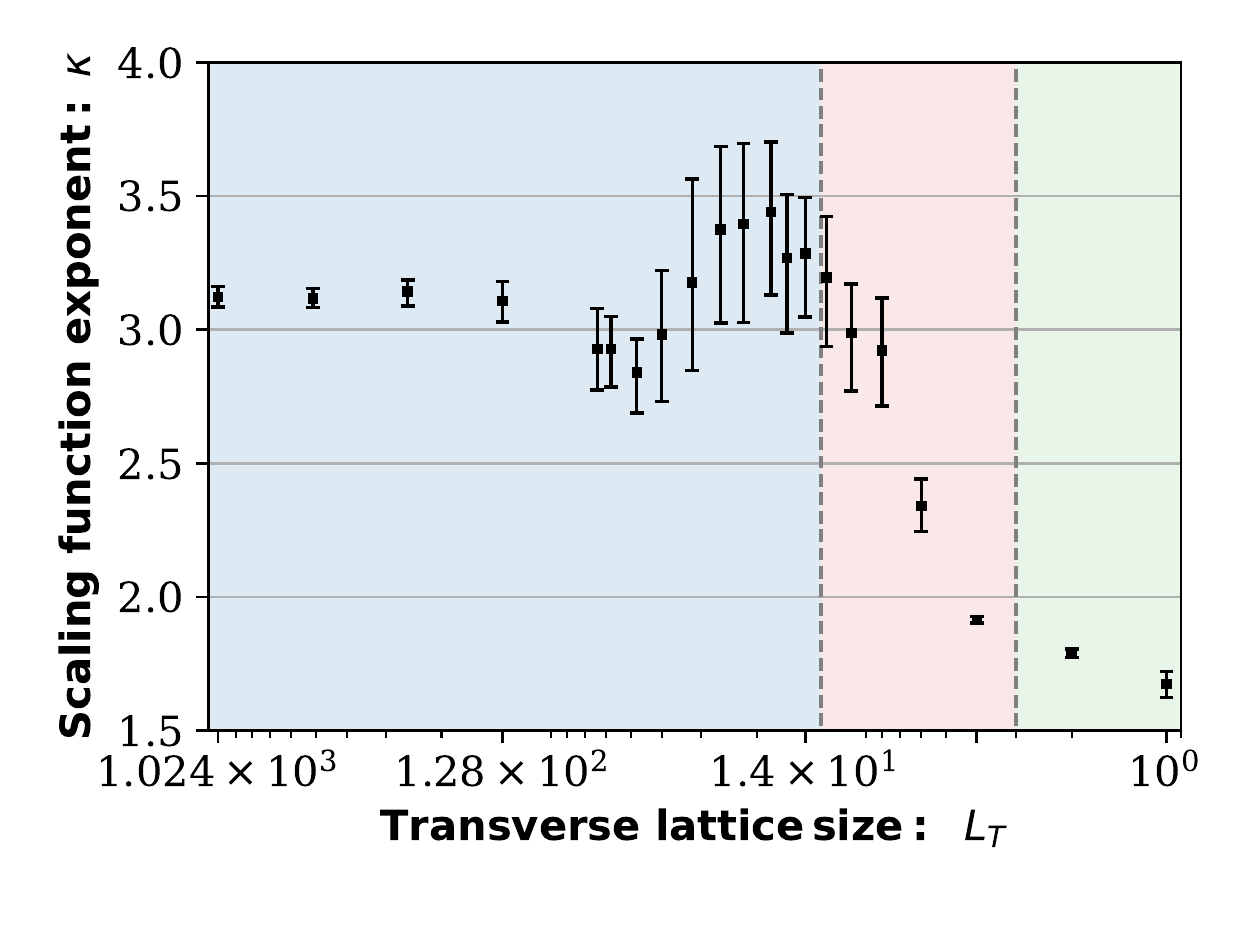}
		\vspace{-0.5cm}
		\caption{Scaling function exponent $\kappa$ as a function of the transverse lattice size. One observes a rather smooth transition in the crossover regime  from two to one spatial dimension.  The dashed lines indicate the boundaries between the three parameter regimes at $L_T \approx 2\pi/Q$ and $L_T \approx 2\pi/\Lambda$.}
		\label{fig: power-law}
	\end{figure}

	As the scaling function exponent $\kappa$ is a universal quantity, it is again interesting to see how it depends on the transverse lattice size. Since the exponent $\kappa$ does not change under rescaling, we can perform a fit according to Eq.~\eqref{eq: fit func} to the original distribution rather then the scaling function and therefore obtain $\kappa$ for each $L_T$ separately. To compute $\kappa$ and its uncertainties as a function of $L_T$, for each lattice size we compute the average over time in the same way as for the other scaling exponents. The results are shown in Fig.~\ref{fig: power-law}. One observes a seemingly smooth transition when going from two dimensions with $\kappa\approx3$ to one dimension with $\kappa\approx2$. The uncertainties of the average again grow relatively large for lattices around $L_T=14$, as the value of $\kappa$ strongly oscillates around the average as a function of time. This is in line with the large uncertainties of the other scaling exponents in that range and may hint at more complicated behavior not captured by our fitting procedure.
	
	
	\section{Discussion and conclusions}
	
	Our results clearly demonstrate the change of the nonequilibrium exponents and scaling functions from two to one spatial dimension as the transverse lattice size is reduced. The correlation exponent $\beta$ turns out to be always close to one-half until it rather abruptly drops to zero for the one-dimensional system. This is consistent with the observed insensitivity of $\beta$ to changes in the dimensionality of space by comparison also to $d=3$ systems~\cite{orioli15}. One-dimensional systems are special in the sense that the phase space for typical (two-to-two) scattering processes between excitations is severely restricted by energy and momentum conservation such that scaling exponents vanish identically. However, we find that even for $L_T \ll L_L$ the vacuum modes turn out to modify the effective interactions for the quasi one-dimensional systems significantly, such that a non-zero $\beta$ can be observed up to the smallest values of $L_T$ for which we properly incorporate the quantum-half in our initial conditions. This is consistent with the experimental observations that quasi one-dimensional setups yield non-zero values for $\beta$. While for the spin-1 system of Ref.~\cite{ober18} a universal value consistent with our result is found, the measurements for the Bose gas of Ref.~\cite{schmied18} yields a significantly smaller value for $\beta$. This indicates that an explanation of the latter requires further ingredients, such as a modified $\beta$ from soliton dynamics~\cite{schmied18}.       
	
	In contrast to $\beta$, our results for the occupation exponent $\alpha$ indicates significant variations for small $L_T$. Since number conservation constrains $\alpha = d \beta$, for constant $\beta$ in $d > 1$ the observed change from $\alpha \simeq 2 \beta $ for $L_T \simeq L_L$ to $\alpha \simeq \beta$ for $L_T \ll L_L$ indeed reflects the effective change in dimensionality. In fact, both quasi one-dimensional experiments of Ref.~\cite{ober18} and Ref.~\cite{schmied18} are consistent with $\alpha \simeq \beta$. 
	
	Remarkably, our findings for the finite-size exponent $\zeta$ and, in particular, the scaling function exponent $\kappa$ clearly indicate a relatively smooth crossover behavior as $L_T$ becomes small. Following the analytical approximate results of Refs.~\cite{PRA99,Walz:2017ffj} from effective kinetic theory, one expects $\kappa \simeq d + 1$ which is consistent with our simulations for the two-dimensional geometry with $L_T \simeq L_L$. For small enough $L_T$ we find rather continuously decreasing values for $\kappa$. For the smallest value of $L_T$ for which we properly incorporate the quantum-half in our initial conditions we find $\kappa$ to be close to two, which accordingly would be consistent with an interpretation as quasi one-dimensional dynamics.  
	
	The fact that several characteristic quantities exhibit a dimensional crossover by varying the finite-size geometry opens up very interesting possibilities also for further experimental studies, in particular, in stronger coupling regimes where our employed theoretical approximation scheme fails. It would be striking if important aspects of the universal behavior of anisotropic systems far from equilibrium can be established to be classified in terms of fractional dimensions.
	
	\acknowledgements
	We thank Sebastian Erne and J\"org Schmiedmayer for discussions.
	This work is supported by the Deutsche Forschungsgemeinschaft (DFG, German Research Foundation) -- Project-ID 27381115 -- SFB 1225 ISOQUANT.
	T.V.Z.'s work is supported by the Simons Collaboration on Ultra-Quantum Matter, which is a grant from the Simons Foundation (651440, P.Z.).

	\appendix
	\section{Self-similarity fitting procedure}
	\label{appendix: fitting procedure}
	To calculate the values of the scaling exponents given in sections \ref{sec:2D} and \ref{sec:crossover} we apply a fitting procedure to the numerically calculated distribution function. The result of this procedure are the scaling exponents as a function of a reference time $t_0$, as shown, e.g., in Fig. \ref{fig: exponents_isotropic}. From these values we then calculate an average as described in the main text. For the isotropic system with $L_T = L_L$ the procedure is exactly the same as described in~\cite{orioli15}.  In the following we therefore only shortly recapitulate the isotropic case and and then explain how we adapt it to systems with $L_T \ll L_L$, including finite-size scaling.
	
	For the isotropic system self-similarity of the distribution function is described by the scaling form \mbox{$f(t,p)=(t/t_0)^\alpha f_S\qty((t/t_0)^\beta p)$}, with $p = |\textbf{p}|$ and a reference time $t_0$. To find the exponents $\alpha$ and $\beta$, we define the rescaled distribution
	\begin{equation}
		f_\text{resc.}(t, p)=(t/t_0)^{-\alpha}f(t, (t/t_0)^{-\beta}p).
	\end{equation}
	Perfect self-similarity implies \mbox{$f_{\text{resc.}}(t, p)-f(t_0,p)=0$}, so in order to find the best estimates for the exponents we minimize deviations from this behavour for the infrared momentum modes. To this end, we define the function 
	\begin{align}
		\label{eq: chisqr}
		&\chi^2(\alpha, \beta)= \nonumber\\
		&\frac{1}{N_\text{comp.}}\sum_{i=1}^{N_\text{comp.}}\frac{\int d(\log{p}) \qty[\qty(f_\text{resc.}(t_k,p)-f(t_0,p))/f(t_0,p)]^2 }{\int d\qty(\log{p})},
	\end{align}
	where $t_k$ are the times considered for the fit and $N_\text{comp.}$ are the number of times we compare. We integrate over $d(\log{p})$ to give a stronger weight to the infrared modes. To calculate $\chi^2$ we linearly interpolate $f_\text{resc.}(t_k, p)$ so that the momenta coincide with the ones of $f(t_0, p)$. The estimates for the scaling exponents $\alpha(t_0)$ and $\beta(t_0)$ are defined as the values that minimize $\chi^2$.
	
	To estimate the error of the scaling exponents for a given $t_0$ we define the likelihood function\begin{equation}
		W(\alpha,\beta)=\frac{1}{\mathcal{N}}\exp\qty(-\frac{\chi^2(\alpha,\beta)}{2\chi^2(\alpha(t_0), \beta(t_0))}),
	\end{equation}
	with the normalization $\mathcal{N}$ so that $\int d\alpha\, d\beta\, W(\alpha,\beta)=1$. We then approximate the marginal likelihood function $W(\alpha)=\int d\beta\, W(\alpha, \beta)$ with a Gaussian and take the standard deviation of the Gaussian $\sigma_\alpha$ as the estimate for the error of $\alpha(t_0)$. Analogously we obtain an estimate for the error of $\beta(t_0)$. These errors are depicted as error bars in Fig. \ref{fig: exponents time}.

	For systems with $L_T \ll L_L$ the distribution function explicitly depends on the transverse size [see Eq.~\eqref{eq:finitesizescaling}] such that it is necessary to perform finite-size scaling. In this case we define the rescaled distribution as 
	\begin{equation}
		f_{0, \text{resc.}}(t, p_L, L_{T,0})=(t/t_0)^{-\alpha}f_0(t, (t/t_0)^{-\beta}p_L, (t/t_0)^{\zeta} L_{T,0}).
	\end{equation}
	The corresponding definition of $\chi^2(\alpha,\beta,\zeta)$ is then the same as in ~\ref{eq: chisqr}, but with the $f_\text{resc.}$ and $f(t_0, p)$ replaced by $f_{0, \text{resc.}}$ and $f_0(t_0, p_L, L_{T,0})$, respectively. 
	The rescaled distribution involves distributions for different times and transverse sizes, which are chosen according to Eq.~\eqref{eq: times_lengths_relation} as described in the main text. The minimization for $\alpha$ and $\beta$ is done for a fixed $\zeta$ with the SciPy library for python~\cite{scipy2020}. We repeat this for the relevant values of $\zeta$ on a grid with a spacing of $\Delta\zeta=0.002$ and select the exponents with the minimal $\chi^2$ value as our estimates for $\alpha$, $\beta$ and $\zeta$. Here error estimation via the likelyhood function as before is not applicable, as for fixed $\alpha$ and $\beta$ the $\chi^2$-function has many local minima, and approximating the resulting marginal likelihood functions, e.g.~$W(\zeta)$, with a Gaussian does not produce accurate results. Instead we only calculate uncertainties for the average value the same way as for the isotropic system as explained in Sec.~\ref{sec:2D}. These averages and their uncertainties are then the scaling exponents depicted in Fig. \ref{fig: exponents final}.
	
	When including lattices from different parameter regimes in the fitting procedure, e.g.\ for \mbox{$L_{T, i}=8, 10, 12, 14$}, we have not found scaling exponents that produce a good fit which we attribute to a missing normalization factor. Since the amplitudes $A$ and $B$ of the scaling function [see Eq.~\eqref{eq: fit func}] are not universal and since different initial conditions in the different parameter regimes lead to significant changes in the energy density also the amplitudes change significantly. We have not found a suitable normalization that removes these differences. Additionally, we find that the scaling exponents change the strongest at the boundaries between the two regimes, which also causes problems in the convergence of the fitting procedure. Therefore, we have excluded the corresponding points for $L_{T,0} = 8, 10, 12$ in Fig. \ref{fig: exponents final}.
	\section{Time evolution of scaling exponents}
	\label{appendix: exponents time crossover}
	\begin{figure}[h]
		\includegraphics[width=\columnwidth]{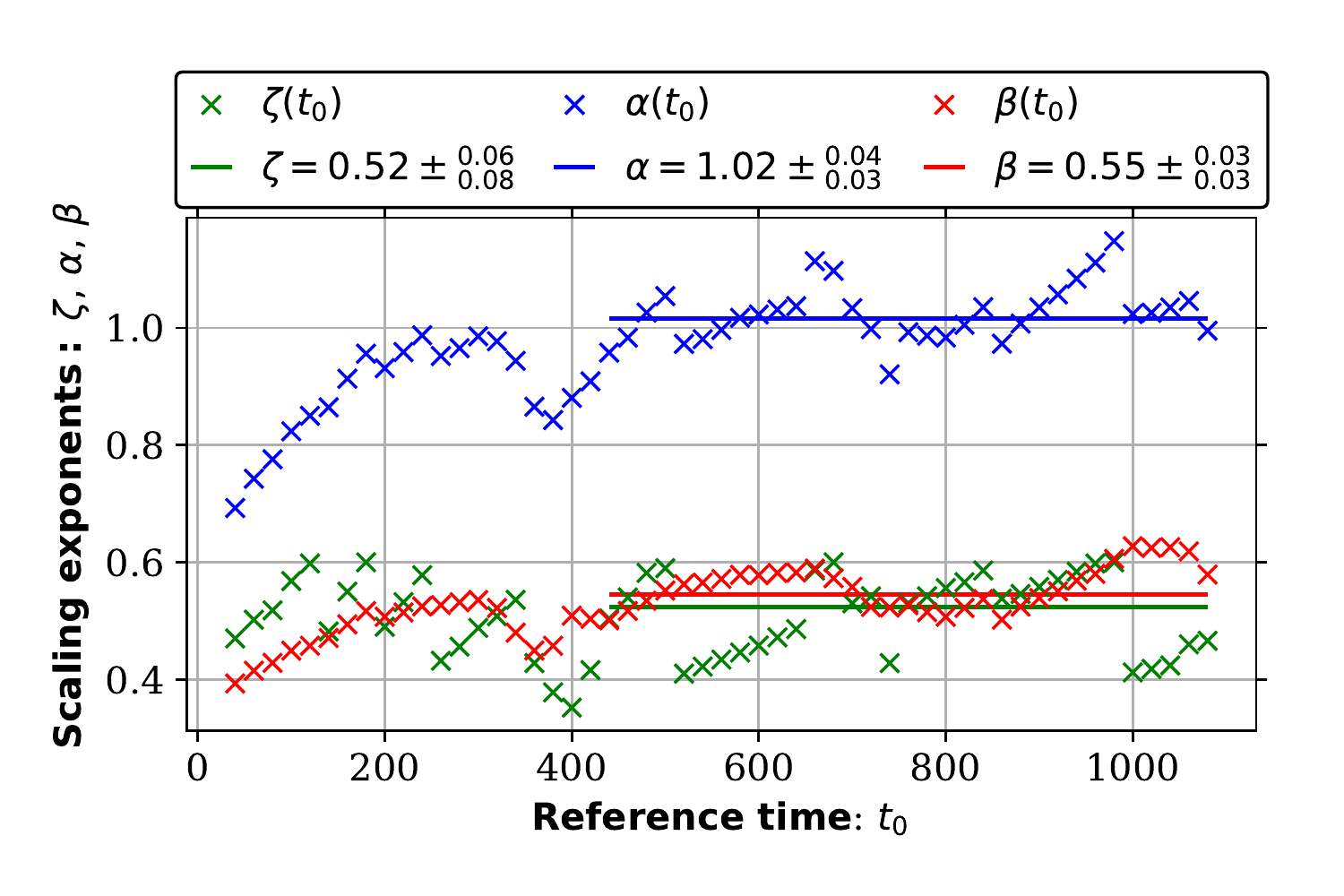}
		\includegraphics[width=\columnwidth]{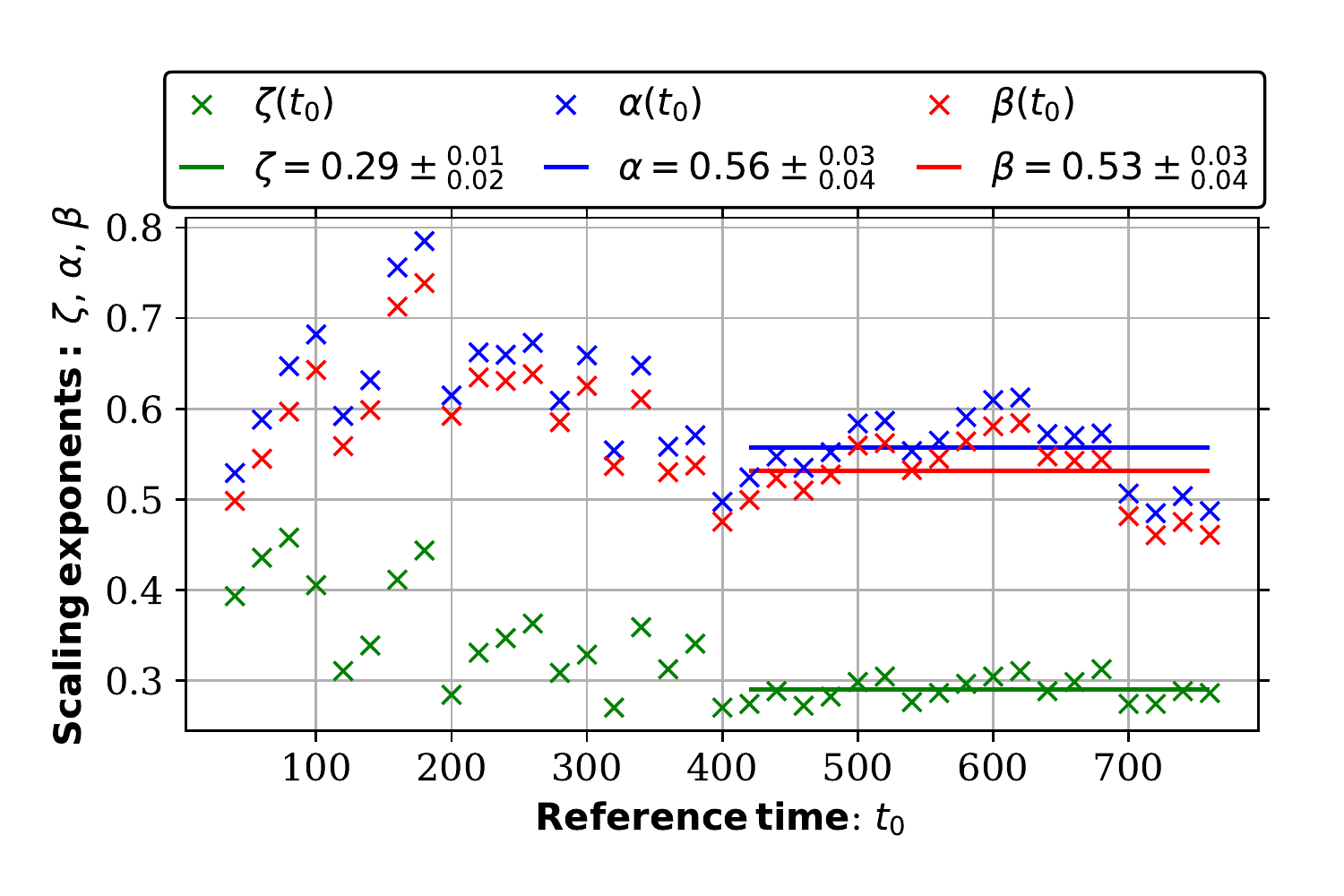}
		\caption{Scaling exponents as a function of reference time $t_0$ for $L_{T,0}=32$ (top) and $L_{T,0}=6$ (bottom). The solid lines illustrate the average values as well as the times included in the averaging.}
		\label{fig: exponents time}
	\end{figure}
	To illustrate how the scaling exponents change with time in the different parameter regimes, we give two exemplifying plots in Fig. \ref{fig: exponents time}, which show the scaling exponents as a function of the reference time $t_0$, as well as their average value, for the two lattice sizes $L_{T,0}=32$ (top) and $L_{T,0}=6$ (bottom). We observe relatively large fluctuations of the exponents, especially for the exponents $\alpha$ and $\zeta$. While the reason for this behavior is not completely clear, we noticed that for different fixed values of $\zeta$ the best fitting exponents $\alpha$ and $\beta$ change significantly. Stronger limits on the range that we sample $\zeta$ from therefore reduce the fluctuations in $\alpha$ and $\beta$ as well, but even for a constant $\zeta$ for all reference times they are still significant. 
	By inspecting the $\chi^2$-function as defined in appendix~\ref{appendix: fitting procedure}, we also noticed the existence of multiple local minima. This explains the rather big jumps of the scaling exponents e.g. at $t_0\approx1000$ for $L_{T,0} = 32$, where the location of the global minimum changes between these local minima.	As mentioned in ~\ref{sec: results exponents} the scaling function exponent $\kappa$  also oscillates quite strongly as a function of time, especially in the range $32\geq L_T\geq14$ (hence the large uncertainties). This indicates that in this regime, the physical behavior might be more complicated than what is captured by our fitting procedure.

\end{document}